\documentclass[a4paper]{article}

\pdfoutput=1
\usepackage{graphicx,wrapfig,lipsum}
\usepackage[export]{adjustbox}
\usepackage{hyperref}
\usepackage{amssymb}
\usepackage{flexisym}

\usepackage[affil-it]{authblk}

\begin{document}

	\title{Dynamic Path Planning and Movement Control in Pedestrian Simulation}
	\author{Fatema T. Johora\thanks{Email: \texttt{ftj14@tu-clausthal.de}}, Philipp Kraus, and J\"org P. M\"uller}
	\affil{Department of Informatics, Clausthal University of Technology, Julius-Albert-Str. 4
		D-38678 Clausthal-Zellerfeld, Germany}	
	
	\maketitle

	\begin{abstract}
		
		Modeling and simulation of pedestrian behavior is used in applications such as planning large buildings, disaster management, or urban planning. Realistically simulating pedestrian behavior is challenging, due to the complexity of individual behavior as well as
		the complexity of interactions of pedestrians with each other and with
		the environment. This work-in-progress paper addresses the tactical (path planning) and the operational level (movement control) of pedestrian simulation from the perspective of multiagent-based modeling. We propose (1) an novel extension of the JPS routing algorithm for tactical planning, and  (2) an architecture how path planning can be integrated with a social-force based movement control. The architecture is inspired by layered architectures for robot planning and control. We validate correctness and efficiency of our approach through simulation runs.
		
	\end{abstract}
	
	\section{Introduction}
\label{sec:Introduction}
Over the past two decades, pedestrian simulation has received considerable attention. Modeling and analysing  the behavior of individual pedestrians provides insights for solving challenging problems such as optimizing building layouts to assure crowd safety. Also, fueled by the paradigm of multiagent-based modeling and by ever increasing computing power, recently we see a trend from macroscopic to microscopic approaches of modeling pedestrians. Crociani et al.~classify the overall behavior of pedestrians in three levels~\cite{ped:behavior}:
\begin{itemize}
	\item At the {\em strategic} level, pedestrians maintain abstract
	plans that motivate their overall behaviour.
	\item At the {\em tactical} level, pedestrians organize their activities, choose activity areas and plan routes to reach their destinations.
	\item Pedestrians' physical movements are handled at the {\em operational} level. 	
\end{itemize}
A comprehensive model of complex pedestrian behavior needs to address each level as well as the interactions
among the levels. The literature on pedestrian simulation contains numerous approaches to handle these levels individually, but there is not much research in integrating these levels. Well-known hierarchical approaches from robot planning (e.g.~\cite{Gat92}) and hybrid agent architectures \cite{Mueller96} have hardly been applied to pedestrian modeling and simulation, possibly due to their computational complexity and limited scalability to large numbers of pedestrians. Our research aims to address this gap to a certain extent. This work-in-progress paper provides a first small step towards applying cognitive layered control architectures to pedestrian simulation. In this paper\footnote{The level of detail in this paper is restricted by space limitations. The master's thesis of the first author \cite{Johora2017} provides a detailed account of the model described in this paper. Accessible at \url{https://goo.gl/rmIBik}.} we focus on tactical (i.e. path planning) and operational (i.e., movement control) pedestrian behavior and their interplay. 
We propose and develop a model to dynamically plan routes for pedestrians and also control their movements during plan execution. We propose a conceptual integration architecture. For path planning, we propose an efficient extension of the JPS algorithm;  for movement control, a social force model is used.  For a proof of concept, we provide a  simple initial implementation of the architecture to integrate path planning and movement control. 

The structure of this paper is as follows: Several works related to this paper are described in Section~\ref{sec:relatedwork}. Section~\ref{sec:requirement} gives an overview of the basic assumptions and requirements underlying our approach. The proposed model is described in
Section~\ref{sec:solution}. In Section~\ref{sec:exp}, we report some experimental findings. We conclude and discuss 
future work in Section~\ref{sec:Conclusion}.
	\section{State of the Art}
\label{sec:relatedwork}
Tactical path planning for pedestrian and controlling their individual motions are two important problems in pedestrian modelling. There is a large body of research in this area; some related approaches are reviewed in this section.  
The concept of steering behaviors, to model realistic movements of an individual character within an environment was introduced by Reynolds~\cite{steer}. This
model proposes different behavioral characteristics such as goal seeking or collision avoidance. The social force model by Helbing~\cite{force} is a physical model describes the movement
of pedestrians by combining different types of attractive and repulsive forces.

In \cite{robot}, Li and Chou propose an approach to plan the motion of robots in crowded environment. Rymill and Dodgson proposed an approach to model
the motion behavior of crowds in real-time~\cite{psycho}. However, this work mainly focuses on the visual perception and collision avoidance techniques of humans.
A model that can plan path for crowds using A* and also supports collision avoidance for controlling their movements is proposed in~\cite{PC}. Cherif and Chighoub \cite{socio-psycho} present a behavioral model of pedestrian movement based on the pedestrian's socio-psychological state.

It can be noted that most of the state-of-the-art research discussed focusses on motion control. Although, notably, the approach in \cite{PC}
implements both path planning and a part of motion control, however, the authors do not explain how these two levels of behaviour are integrated. However, in research on mobile robot control and agent architectures, there are many concepts for integrating planning with execution, such as the ATLANTIS architecture proposed by Gat~\cite{Gat92}, or the layered {\sc InteRRaP} architecture described in~\cite{Mueller96}. These approaches motivate us to consider the topic of integrating path planning and movement control behaviors of pedestrians.
	\section{Problems and Requirements}
\label{sec:requirement}
Normally, pedestrians plan their path based on the current state of their working environment or sometimes even before entering their working environment. They do not consider the dynamically changing
environment including other pedestrians, during path planning.  However, due to the changes in their local environment such as extreme crowd, they may need to re-plan their path or somehow try to adapt the changes~\cite{Fr16}. 
Our assumption is that pedestrian do not always plan path using straight minimal-length path metric. But we use this metric to design our overall architecture and the algorithms with a motivation that we can start with simple objective functions and, in future work, replace them by more realistic objective functions, without the need of replacing the path planning algorithm.
To simulate the walking behavior of pedestrians, two levels of behaviors namely planning (tactical level) and execution are required to be integrated (operational level). 

To plan the shortest path for pedestrians, a path planning algorithm is needed. A mobility model is also needed to control the movement of pedestrians by satisfying the dynamic aspects of their environment. Finally, a handler is needed to handle the interaction between path planning and movement control.

	\section{Integrated Control Architecture}
\label{sec:solution}
We propose a conceptual architecture that implements pedestrian path planning and movement control and  handles the collaboration between the two levels. This architecture consists of three modules which are described in the following subsections. The \textbf{path planner} (Section~\ref{subsec:path_planning}) provides (deliberative) route planning on a graph or grid model (in this paper, we assume a grid-based representation). The \textbf{motion controller} (cf.~Section~\ref{subsec:motion-ctrl}) handles the operational (reactive) moving decisions of pedestrians. The control flow between these two modules is orchestrated the third component: The  \textbf{transition handler} (Section~\ref{subsec:transition_hdlr}). In the approach described in this paper, we assume that the path planner operates at a high level of abstraction, not taking dynamic aspects of the environment into consideration, while the motion controller accommodates these dynamic aspects. 

\subsection{Path Planner}\label{subsec:path_planning}
For pedestrian path planning, we propose and develop an optimization of the Jump Point Search (JPS) algorithm~\cite{jps}. We start by explaining the basic model underlying JPS. JPS combines an A* algorithm with two sets of rules, i.~e.~\emph{pruning rules} and \emph{jumping rules}:
\paragraph{Pruning Rules:} JPS chooses a single path among many equivalent
paths using the following pruning strategy:

\begin{itemize}
	\item If current node $x$ can be reached through a straight move from its
	parent $p(x)$, then JPS prunes any neighbor of $x$ that
	satisfies the constraint $length(p(x), ...,n) - x) \leq length(p(x),x,n)$
	
	\item If $x$ is a diagonal move away from $p(x)$, JPS prunes any neighbor of
	$x$ that satisfies the constraint $length(p(x), ...,n) - x) < length(p(x),x,n)$	
\end{itemize}

If $x$ does not have any neighbor which is an obstacle, then the
remaining nodes after applying pruning rules are called its natural neighbors.
If any neighbor of $x$ isan  obstacle then any remaining node $n$ after pruning is a forced neighbor of $x$, if (1) $n$ is not a natural neighbor of the current node $x$, and (2)  $length(p(x), x, n)$ $<$ $length(p(x), ..., n) - x)$.

\paragraph{Jumping Rules:} JPS does not expand all non-pruned neighbors of $x$ as its successors. It continues moving in the direction of each non-pruned neighbor until it finds jump points (JPs). Any non-pruned neighbor $y$ of $x$, located in a direction $\vec{d}$, can be a JP if $y$ minimizes the number of steps $k$ such that $y = x+k\vec{d}$ and also satisfies one of the following constraints:
\begin{enumerate}
	\item y is the goal node
	\item y has at least one forced neighbor
	\item If $\vec{d}$ is a diagonal move and $z = y+ k_i\vec{d_i}$ is a jump point from node $y$ by satisfying constraint 1 or 2. Here, $k_i$ is a real number and $\vec{d_i} \in (\vec{d_1}, \vec{d_2})$
\end{enumerate}

\begin{figure}[htbp]
	\begin{center}
	\includegraphics[width=.75\textwidth]{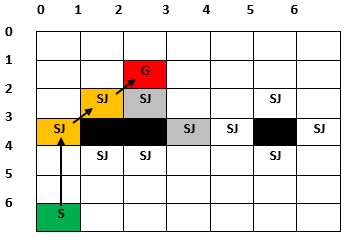}
	\caption{JPS-S path planning returns a sequence of jump points. Orange nodes indicate
		expanded nodes. Nodes with SJ denote pre-calculated
		JPs, gray nodes denote JP successors of expanded nodes.}\label{wrap-fig:1}
\end{center}
\end{figure}

JPS first expands the start node (S) to get all its JP successors, then recursively expands the successors, until it finds the target (G). At each iteration, it chooses one node to expand among all successors, based on the estimated cost to go to G from S via each successor. In a more recent paper, the authors of JPS introduced Jump Point Search Plus (JPS+)~\cite{jps+}, which performs a dynamic (and rather resource-intensive) calculation of jump points.

To increase the efficiency of JPS / JPS+ for larger simulations, we propose an optimization named \emph{JPS with Static Jump Points (JPS-S)}.In JPS-S, we add an offline pre-processing step to JPS. Here, offline preprocessing stands for all the calculations which are performed before starting the actual path finding. Unlike JPS+, JPS-S pre-calculates the  JPs surrounding static obstacles on the grid map; these JPs are called static jump points (SJPs) as they remain  unchanged unless the map changes.
JPS+ pre-calculates the first JPs in eight possible directions for every traversable node on the grid map, which enhances the performance of JPS but also increases the memory usages significantly. Our approach JPS-S speeds up JPS, but with minimized memory usage (see Section~\ref{sec:exp}).

Before starting path planning, JPS-S filters the SJP list using the given start and goal positions to get all SJPs within the search range.
The path searching procedures of JPS and JPS-S algorithms are similar, except that while searching the next JP in diagonal direction, JPS-S is searching recursively in the same direction that JPS does, but, in contrast to JPS, it checks the SJP list before. If it finds any SJP in that direction, it considers it as the successor of the currently expanding node.

\subsection{Motion Controller}\label{subsec:motion-ctrl}
We select the social force model~\cite{force} for controlling the movement of pedestrian. In the social force model, the motion of a pedestrian is conducted by the following set of simple forces that a pedestrian faces at a specific time:
\paragraph{Acceleration:} The model assumes that pedestrians move towards their destination taking the shortest possible way. The pedestrian $\alpha$ reaches his/her destination with a desired velocity $\vec{w}_\alpha(t)$ unless the movement of $\alpha$ is distracted. If his/her velocity deviates to $\vec{v}_\alpha(t)$ then he/she accelerates to achieve $\vec{w}_\alpha(t)$ again in a certain relaxation time $\tau$. This acceleration can be computed by the  formula:
\begin{equation}
\vec{f}_{\alpha}^o(\vec{w}_\alpha(t), \vec{v}_\alpha(t) ) = \frac{\vec{w}_\alpha(t) - \vec{v}_\alpha(t)}{\tau}
\end{equation}
\paragraph{Repulsive Forces from Other Pedestrians and Boundaries:}
\label{repulseothers}
Pedestrians maintain some distance from unknown pedestrians ($\beta$), boundaries or static obstacles (B). These behaviors can be represented by the two  force vectors
\begin{equation}
f_{\alpha\beta}^{soc}(\vec{d}_{\alpha\beta}(t)) = V_{\alpha\beta}^o\exp\bigg[\frac{- \vec{d}_{\alpha\beta}(t)}\sigma\bigg] \vec{n}_{\alpha\beta} F_{\alpha\beta}
\end{equation}
and 
\begin{equation}
\vec{f}_{\alpha B}^{soc}(\vec{d}_{\alpha B}(t)) =  U_{\alpha B}^o\exp\bigg[\frac{- \vec{d}_{\alpha B}(t)}{R}\bigg] \vec{n}_{\alpha B}.
\end{equation}
Here, $V_{\alpha\beta}^o$ and $U_{\alpha B}^o$ denote the interaction strengths and $\sigma$ and $R$ indicates the range of these repulsive interactions. $\vec{d}_{\alpha\beta}(t)$ and $\vec{d}_{\alpha B}(t)$ are the distances from $\alpha$ to $\beta$, and $\alpha$ to $B$ at a specific time. $\vec{n}_{\alpha\beta}$ and $\vec{n}_{\alpha B}$ denote the normalized vectors. $F_{\alpha\beta}$ represents the anisotropic behavior of pedestrian i.e. pedestrians are mostly influenced by the objects which capture within their angle of view.
\paragraph{Attractive Forces towards Point of Interests:}
Pedestrians can be attracted to their point of interests such as street artists. This attractive force $\vec{f}_{\alpha i}^{soc}$ is time-dependent and linearly decreased to zero. It is calculated by the same equation as $\vec{f}_{\alpha\beta}^{soc}$. As compared to repulsive interactions, the interaction range of $\vec{f}_{\alpha i}^{soc}$ is larger and the strength of this interaction is smaller, negative and time-dependent.
\paragraph{Joining Behavior:}
Pedestrians aim to stay together as a group with their family or friends. The joining behavior represents this behavior~\cite{joining}. It is calculated by the formula of the interaction strength $C_{\alpha\beta}$ and the normalized vector $\vec{n}_{\beta\alpha}$. Combination of these forces describes the temporal change  of pedestrian's velocity.
\subsection{Transition Handler}\label{subsec:transition_hdlr}
\begin{figure}[htbp]
	\begin{center}
	\includegraphics[width=.8\textwidth]{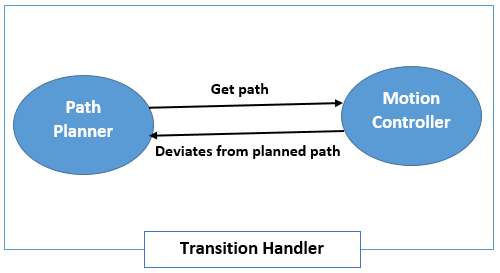}
	\caption{Control architecture}
	\label{fig:chapter01:overview}
	\end{center}
\end{figure}

The control flow between path planner and motion controller can be realized in largely differing ways: In a naive model, planner and  controller may work sequentially. In a more dynamic model, they may be represented as states of a finite state machine, and take control alternatively. In a very sophisticated model, they may run concurrently with a vertical or horizontal flow regulating access to perception and action (see e.g.~\cite{Mueller96}). In this paper, we propose a conceptual control architecture where 
transition handler (TH) component decides when control is shifted between path planner and motion controller. This can be modeled by a simple finite state machine with two states, where only one state has control of the pedestrian at a specific time (see Figure~\ref{fig:chapter01:overview}). State transition is determined based on the currently active state, perceiving input and transition conditions. If the perceptual input trigger any transition condition of the current state then that transition will be conducted. The TH receives a set of jump points specifying the desired path from the path planner. Then, it switches to the controller state; the movement of a pedestrian from its current position to the next jump point is then handled by the motion controller. 

The transition handler will also be responsible for plan monitoring: If during plan execution, pedestrians deviate from their pre-calculated path, the TH will recognize this and shift the locus of control to the path planner state, where the path will be recalculated. For identifying a deviation, we use an algorithm very similar to the path following behavior in Reynolds' steering model~\cite{steer}. It represents a path as a spine and a radius; a spine thus represents a sequence of line segments. If the distance between the next position of the pedestrian and the nearest point on the path is greater than the path radius, then a deviation is detected.
	\section{Evaluation}
\label{sec:exp}
To evaluate our proposed model, we tried to answer the following questions:
\begin{itemize}
	\item Does JPS-S plan optimal-length paths? Does it optimize the JPS algorithm?
	\item Does the social force model accommodate the movement behaviors of pedestrians?
	\item How to integrate the path planning and movement control behaviors
	of pedestrian in a desirable way?
\end{itemize}
To answer these questions, we present an argumentation on the optimality of JPS-S in terms of minimal length and also perform computer
simulations. Simulations are implemented using the Java (version 8) programming language on an Intel$\circledR$ Core\texttrademark i5 processor with 3 GB RAM.

\paragraph{Qualitative Evaluation of Minimality of Paths Found by JPS-S:} 
\label{correctness}
Both JPS and JPS-S algorithms give a sequence of jump points (JPs) as the optimal- length path. The proof that the optimal path between two points (if exists)
in a grid map can be found by only expanding JPs during search, is given in~\cite{jps}. In this paragraph, we argue that the
extension of the pre-processing technique (SJP) with JPS maintains optimality
and completeness of JPS, as follows:

Let $\pi$ = \{ $S$, $JP_1$, $JP_2$, $JP_3$....$JP_n(G)$ \} be the optimal-length path between S and G, which is calculated by JPS. Let further $\pi\textprime$ = \{ $S$, $JP_1\textprime$, $JP_2\textprime$, $JP_3\textprime$....$JP_n\textprime(G)$ \} be the path between S and G, which is calculated by JPS-S. Let length($JP_x,JP_y$) denote the euclidean distance between $JP_x$ and $JP_y$.

Any $JP_x$ = $JP_x\textprime$, if the direction from $JP_{x-1}\textprime$ to $JP_x\textprime$ is straight. If the direction from $JP_{x-1}\textprime$ to $JP_x\textprime$ is diagonal then for d = length($JP_{x-1}$,$JP_x$) and $d\textprime$ = length($JP_{x-1}\textprime$, $JP_x\textprime$), $ d \leq d\textprime$ holds. For each $JP_x$ and $JP_x\textprime$, if $d\textprime = d$ then the number of JPs in $\pi$ and $\pi\textprime$ are equal. Otherwise, the number of JPs in $\pi\textprime$ is smaller than the number of JPs in $\pi$ because length($JP_{x-1}\textprime$,$JP_x\textprime$) = length($JP_{x-1}$,$JP_x$) + length($JP_x$,$JP_{x+1}$). The length of $\pi$ and $\pi\textprime$ is always the same.

JPS-S always maintains the basic properties of JPS: the pruning rules, the jumping rules, all moves to go from one JP to another adjacent JP always involve traveling in the same direction, and the line between two adjacent JPs does not collide with any obstacle.

\paragraph{Performance Test:}

\begin{figure}[htbp]
	\begin{center}
	\includegraphics[width=.8\textwidth]{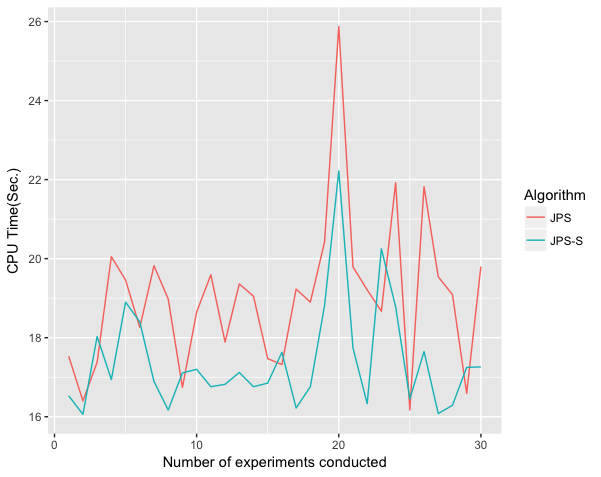}
	\caption{JPS-S outperforms JPS with respect to time efficiency (CPU time).}
	\label{fig:compare}
	\end{center}
\end{figure}

A grid-based map with 250 $\times$ 250 cells is used as the environment for
performing path planning using JPS-S. We measure performance
in terms of the relative improvement to the needed time to solve a given
problem. We perform tests on both the JPS and JPS-S algorithms, to find the optimal path for 10000 pedestrians in parallel at a time. This test is performed 30 times. Figure~\ref{fig:compare} shows the results of this test. On average, the test
takes approximately 9\% less computation time for JPS-S. In the best case, the test takes over 16\% extra time for JPS than for JPS-S.

We use the following simulation benchmarks to test the efficiency of the social force model and all of these simulations terminates successfully as each pedestrian reaches their goal point.

\begin{description}
	\item[Narrow Walkway:] Two oppositely directed groups of 50 pedestrians walking through a narrow walkway to reach their goal by avoiding collisions.
	\item[Narrow Passage:] A group of 100 pedestrians are trying to pass a narrow door to reach their goal by avoiding collisions with others. 
	\item[Path Following:] Two groups of 100 pedestrians are trying to reach similar goal by moving along their planned path and avoiding collisions with others. 
\end{description}

\paragraph{Interpretation of Results:}
The results of the above small experiments show that the
performance of JPS-S is better than that of JPS in terms of
needed time; also in the benchmark scenarios, the social force model suitably accommodates the movement behavior of pedestrians. The qualitative evaluation of the optimality of JPS-S is also given in page~\pageref{correctness}.
Therefore, the first two questions which are described in the beginning of this section are answered.

JPS-S is time efficient and the social force model efficiently enables pedestrians to move along their planned path which is validated from the result of path following benchmark. Hence, we conclude that our architecture integrates the path planning and movement control behaviors
of pedestrian in a correct and efficient way. 

	\section{Conclusion and Outlook}\label{sec:Conclusion}
This paper reports work-in-progress towards a multiagent-based model for pedestrian simulation that integrates tactical and operational behaviour. The contribution of this paper is threefold: 
First, we propose a conceptual control architecture to integrate pedestrian path planning and movement control. Second, we propose an optimization JPS-S of the well-known JPS algorithm for shortest path planning. We show that JPS-S maintains the optimality properties of JPS, and we compare it experimentally against JPS and JPS+, showing its advantages in performance and scalability. Third,  we describe a movement control algorithm based on the social force model~\cite{force}, and implemented a proof-of-concept of the overall system combining JPS-S with the social force model. Simulation results from this proof-of-concept indicate the feasibility and validity of the overall model, but also some limitations of the approach. 

So far we only implemented the naive integration architecture without re-planning. Also, the fact that the JPS-S algorithm places the static jump points close to static obstacles may not work well with the social force model which tends to steers pedestrians away from obstacles. Dedicated experiments are needed to explore this relationship. 
Also so far, the simulation is implemented in plain Java. We are working to port it to the LightJason BDI agent platform~\cite{Kraus+2016eumas} for better scalability. Finally, so far we have not yet compared the output of our model with benchmarks of real pedestrian behaviours. This is another necessary task for future work.

\bibliography{main}
\bibliographystyle{plain}

\end{document}